\documentclass[aip,apm,amsmath,amssymb,preprint]{revtex4-2}
\usepackage{graphicx}% Include figure files

\usepackage[utf8]{inputenc}

\begin{document}

%\preprint{}

\title{Origin of oscillatory structures in the magnetothermal conductivity of the putative Kitaev magnet $\alpha$-RuCl$_3$ }



\author{J.A.N. Bruin}
%\thanks{}
\author{R.R. Claus}
\author{Y. Matsumoto}
\author{J. Nuss}
\affiliation{Max Planck Institute for Solid State Research, Heisenbergstraße 1, 70569 Stuttgart, Germany}\author{S. Laha}
\altaffiliation[Present address: ]{Department of Humanities and Sciences, Indian Institute of Petroleum and Energy, Visakhapatnam - 530003, Andhra Pradesh, India}
\affiliation{Max Planck Institute for Solid State Research, Heisenbergstraße 1, 70569 Stuttgart, Germany}
\author{B.V. Lotsch}
\affiliation{Max Planck Institute for Solid State Research, Heisenbergstraße 1, 70569 Stuttgart, Germany}
\affiliation{Department of Chemistry, University of Munich (LMU), Butenandtstraße 5-13, 81377 Munich, Germany}
\author{N. Kurita}
\affiliation{Department of Physics, Tokyo Institute of Technology, 2-12-1 Oh-Okayama, Meguro-ku, Tokyo 152-8551, Japan}
\author{H. Tanaka}
\affiliation{Department of Physics, Tokyo Institute of Technology, 2-12-1 Oh-Okayama, Meguro-ku, Tokyo 152-8551, Japan}
\author{H. Takagi}
\affiliation{Max Planck Institute for Solid State Research, Heisenbergstraße 1, 70569 Stuttgart, Germany}
\affiliation{Institute for Functional Matter and Quantum Technologies, University of Stuttgart, Pfaffenwaldring 57, 70550 Stuttgart, Germany}
\affiliation{Department of Physics, University of Tokyo, 7-3-1 Hongo, Bunkyo-ku, Tokyo 113-0033, Japan}

\date{\today}

\begin{abstract}
The layered honeycomb magnet $\alpha$-RuCl$_3$ has been suggested to exhibit a field-induced quantum spin liquid state, in which the reported large thermal Hall effect close to the half-quantized value still remains a subject of debate. Recently, oscillatory structures of the magnetothermal conductivity were reported and interpreted as quantum oscillations of charge-neutral particles. To investigate the origin of these oscillatory structures, we performed a comprehensive measurement of the in-plane magnetothermal conductivity $\kappa(H)$ down to low temperature (100 mK), as well as magnetization $M$, for single crystals grown by two different techniques: Bridgman and chemical vapor transport. The results show a series of dips in $\kappa(H)$ and peaks in the field derivative of $M$ located at the same fields independent of the growth method. We argue that these structures originate from field-induced phase transitions rather than quantum oscillations. The positions of several of these features are temperature-dependent and connected to the magnetic phase transitions in zero field: the main transition at 7 K and weaker additional transitions which likely arise from secondary phases at 10 K and 13 K. In contrast to what is expected for quantum oscillations, the magnitude of the structure in $\kappa(H)$ is smaller for the higher conductivity crystal and decreases rapidly upon cooling below 1 K.
\end{abstract}

\maketitle %\maketitle must follow title, authors, abstract

\section{Introduction}

Kitaev quantum spin liquids have an exactly solvable ground state in which the excitations are Majorana fermions \cite{Kitaev2006,Jackeli2009}, and the $J_{\textrm{eff}}$ = $\frac{1}{2}$ layered honeycomb magnet $\alpha$-RuCl$_3$ has emerged as one of the prominent material candidates\cite{Plumb2014,Trebst2017,Takagi2019}. In the absence of an applied magnetic field, the ground state has zig-zag antiferromagnetic order ($T_\textrm{N}$ $\approx$ 7 K), but upon applying an in-plane magnetic field the ordered state is suppressed ($H_{\text{C2}} \approx$ 7 T) and a quantum paramagnetic regime is accessed \cite{Johnson2015,Kubota2015,Sears2015}. In this field-induced state, a thermal Hall conductivity close to the half-quantized value was reported and discussed as a response of a topological edge state of Majorana fermions \cite{Kasahara2018a,Nasu2017}. Although that result and interpretation are the subject of ongoing debate \cite{Kasahara2018a,Yamashita2020,Yokoi2021,Yamashita2020,Czajka2021a,Lefrancois2021,Bruin2022,Czajka2022}, the field-induced state clearly exhibits exotic behavior.

Another unusual property of the field-induced state was reported recently by Czajka \textit{et al.}\cite{Czajka2021a}, who observed a sequence of minima in the magnetic field dependence of thermal conductivity $\kappa$ and interpreted them as quantum oscillations originating from a Fermi surface of neutral quasiparticles. The amplitudes of these oscillations were shown to grow upon cooling down to $\sim$1 K, in line with the expected Lifshitz-Kosevich behavior. The present authors, however, have pointed out that two of the reported oscillation minima coincide with known magnetic phase transitions, the sequential collapse of zig-zag order at  $H_{\text{C1}}$ and $H_{\text{C2}}$, and that amplitudes of these oscillations collapse upon cooling below 1 K\cite{Bruin2022}. These observations contradict expectation for conventional quantum oscillations. The origin of the oscillatory features remains to be examined carefully by more comprehensive experiments over a wide range of temperatures and fields. We note that, while the measurements by Czajka \textit{et al.} were conducted using single crystals grown by a chemical vapor transport (CVT) method, the measurements by the present authors were done on a Bridgman-grown sample (the same batch of single crystals as those used in the original report of half-quantization \cite{Kasahara2018a}). The possibility of a sample dependence must also be considered when addressing the controversial issue of oscillations.

In this study we focus on the question of the origin of the oscillatory structures in the magnetic field dependence of thermal conductivity $\kappa(H)$ of $\alpha$-RuCl$_3$.  We present detailed measurements of magnetothermal conductivity along the $a$-axis and magnetization with the in-plane field oriented along either the crystal $a$ or $b$ axes, down to the low temperature of 100 mK for single crystals grown by both the Bridgman and the CVT methods. All the experimental results point to magnetic phase transitions as the origin of the oscillatory structures, independent of the crystal growth method.

\section{Methods}

Two single crystal pieces grown by a CVT- and a Bridgman technique were measured. The CVT-grown crystal was sourced from the same batch as reported in Ref.~\onlinecite{Suzuki2021}. The Bridgman-grown crystal was the same as reported in Ref.~\onlinecite{Bruin2022}. They were both plate-like and similar in size, with approximate dimensions 2 mm x 1 mm x 0.02 mm (length x width x thickness). The crystal structures and their orientations were determined by single-crystal x-ray diffraction at room temperature using a SMART-APEX-II CCD X-ray diffractometer (Bruker AXS, Karlsruhe, Germany) with graphite monochromated MoK$\alpha$ radiation. The Bridgman sample was best fitted to a C2/m monoclinic unit cell with lattice parameters a = 6.041(4) \AA, b = 10.416(8) \AA, c = 6.088(4) \AA, $\beta$ = 108.54(2)$^{\circ}$, and the CVT sample was determined to be trigonal P3$_1$12 with lattice parameters a = b = 6.012(3) \AA, c = 17.27(1) \AA. Lattice parameters were refined with the SAINT subprogram in the Bruker Suite software package\cite{Bruker2019}, using 448 and 776 reflections, respectively. Despite the differences in the room temperature structures of these two samples, we will see that their low temperature conductivities and magnetic properties are similar to each other.

Thermal conductivity was measured using a one heater- two thermometer technique on a dilution refrigerator. RuO$_2$ chip thermometers, calibrated in-situ against a field-calibrated RuO$_2$ reference, were used to determine the temperature gradients. To measure the magnetic field dependence of thermal conductivity, the sample was allowed to thermalize fully at each field before every measurement so that true steady state conditions were achieved. Complementary continuous field sweeps were performed in some cases to examine the detailed field dependence, and these data were cross-checked against steady-state field sweeps to verify consistency. All measurements of $\kappa$ were performed with heat applied along the crystal $a$-axis. The Bridgman sample $\kappa$ data with $H \parallel a$ plotted in Figs. \ref{fig1}b,c and \ref{fig2}b,d are reproduced from Ref.~\onlinecite{Bruin2022}. For the CVT crystal, magnetization data before and after mechanical bending are presented, and all thermal conductivity data for that crystal were measured before it was bent.

The crystal $a$ and $b$ axes are defined as being perpendicular and parallel to the Ru-Ru bond directions, respectively. In the case of the monoclinic C2/m structure of the Bridgman crystal, the $a$-axis is uniquely determined as the direction of the layer-to-layer Ru honeycomb shift. In a monoclinic unit cell, the honeycomb axes perpendicular and parallel to Ru-Ru bonds other than the $a$ and $b$ axes are distinct, and we denote these as $a$*,$a$** and $b$*,$b$**, respectively (Fig.~\ref{fig4}a). In the case of the trigonal CVT crystal, $a$ was defined as the direction of the longest sample dimension. Magnetization measurements were performed on both samples using a Physical Properties Measurement System vibrating sample magnetometer option (Quantum Design, USA) along all 6 named crystal axes.

\section{Results and discussion}

\begin{figure*}
\includegraphics[width=0.8\textwidth, angle=0]{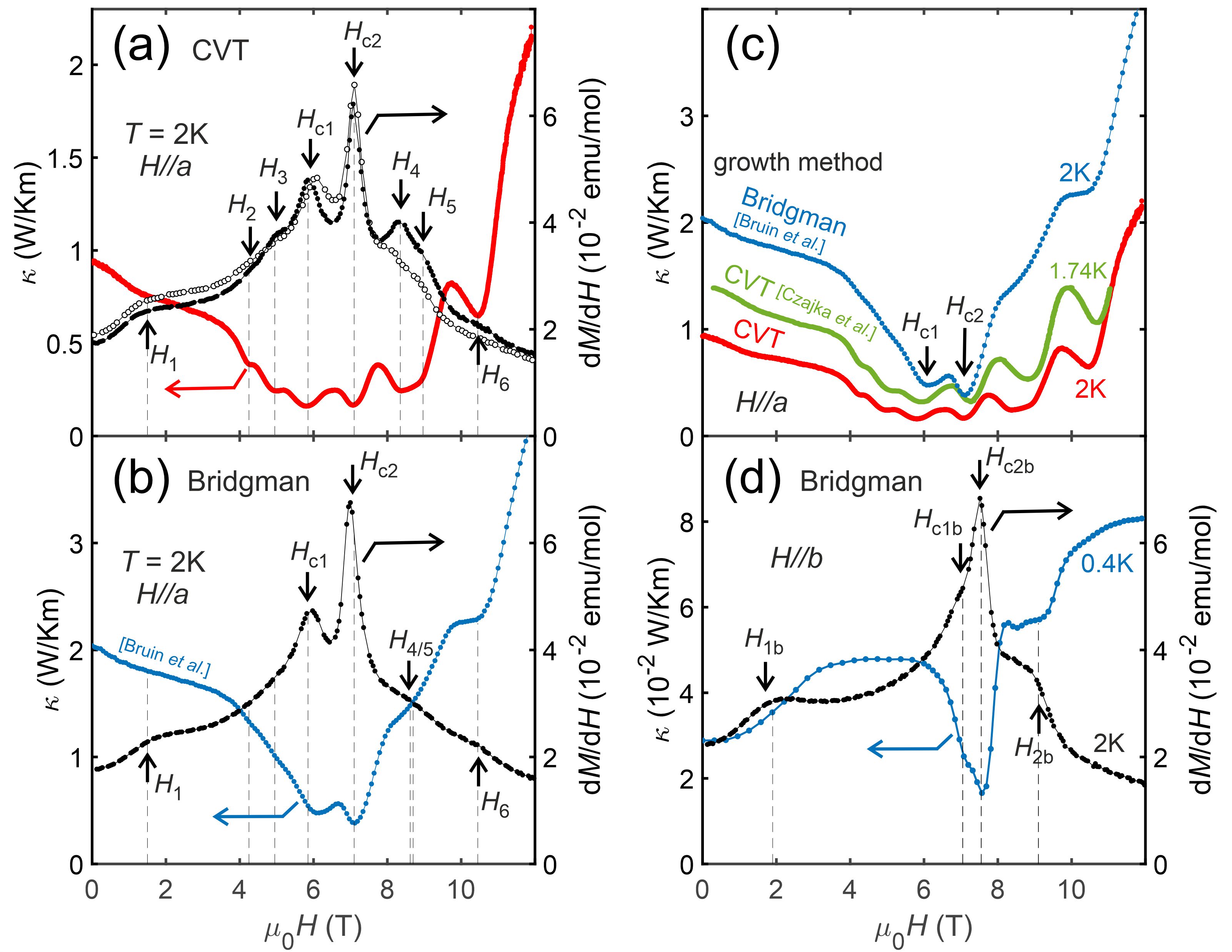}
\caption{\label{fig1} (a) Isotherms of thermal conductivity $\kappa$ (red) and d$M$/d$H$ (black) with $H \parallel a$ at $T$=2 K for the same CVT-grown sample. Open and closed black markers denote d$M$/d$H$ before and after the sample was lightly bent with tweezers, respectively. Several prominent peaks are enhanced by bending. Dotted lines indicate the locations of minima or kinks in $\kappa(H)$ and arrows indicate the coincident peaks or shoulder features in d$M$/d$H$. (b) Isotherms of thermal conductivity $\kappa$ (blue) and d$M$/d$H$ (black) with $H \parallel a$ at $T$=2 K for the same Bridgman-grown sample. Dotted lines indicate the same fields as in panel a. Underlying $\kappa(H)$ data were reproduced from Ref.~\onlinecite{Bruin2022} (c) Isotherms of thermal conductivity around 2 K with $H \parallel a$, comparing the Bridgman (blue) and CVT-grown (red) samples, as well as the result from Ref.~\onlinecite{Czajka2021a} (green). All principal features are reproduced, and the similarity of the two CVT-grown samples is especially striking. (d) Isotherms of thermal conductivity $\kappa$ (blue) and d$M$/d$H$ (black) with $H \parallel b$ for the Bridgman sample. Dotted lines indicate the locations of minima or kinks in $\kappa(H)$ and arrows indicate the coincident peaks or shoulder features in d$M$/d$H$.}
\end{figure*}

The magnetothermal conductivity displays rich structures in both the Bridgman and the CVT crystals, for magnetic field applied along both $a$ and $b$ axes as shown in Figs.~\ref{fig1}a, \ref{fig1}b and \ref{fig1}d. In all the studied crystals and field configurations, the observed minima or kinks in $\kappa(H)$ coincide with the magnetic fields where the field derivative of magnetization d$M$/d$H$ has a peak or shoulder, including at the magnetic phase transitions associated with the collapse of zig-zag order at $H_{\text{C1}}$ and $H_{\text{C2}}$, which already suggests a link between magnetic transitions and the structures in $\kappa(H)$. In Fig.~\ref{fig1}c, the magnetothermal conductivity with $H \parallel a$ at 2 K for the Bridgman and the CVT crystals are compared together with the data from Ref.~\onlinecite{Czajka2021a} at 1.75 K. The absolute magnitude of the conductivity of our CVT-grown sample, as well as the sample in Ref.~\onlinecite{Czajka2021a}, is lower than that of the Bridgman-grown sample, which suggests more disorder in the former. However, the structures are more prominent for the CVT-grown samples, so the degree of disorder correlates with the magnitude of oscillatory features, which goes against expectation for canonical quantum oscillations in metals; the cleaner the sample is, the larger the amplitude of oscillations. The positions of minima occur at nearly the same magnetic fields for all three crystals: at $H_{\text{1}}\sim$1.5 T, $H_{\text{2}}\sim$4.2 T, $H_{\text{3}}\sim$4.8 T, $H_{\text{C1}}\sim$5.8 T, $H_{\text{C2}}\sim$7 T, $H_{\text{4}}\sim$8.3 T, $H_{\text{5}}\sim$9 T and $H_{\text{6}}\sim$10.5 T (Figs.~\ref{fig1}a,b), except that the distinction between $H_{\text{4}}$ and $H_{\text{5}}$ is unresolved for the Bridgman sample. For the Bridgman sample, the features are clearly less prominent in both $\kappa(H)$ and d$M$/d$H$ than in the CVT sample, in particular $H_{\text{2}}$ and $H_{\text{3}}$ are very weak in $\kappa(H)$ and not resolved in d$M$/d$H$. For magnetic field applied along the $b$-axis we resolve fewer features, and their positions are different than for $H \parallel a$ (Fig.~\ref{fig1}d), but the coincidence between structures in $\kappa(H)$ and maxima or shoulders in d$M$/d$H$ persists.

\begin{figure*}
\includegraphics[width=0.8\textwidth, angle=0]{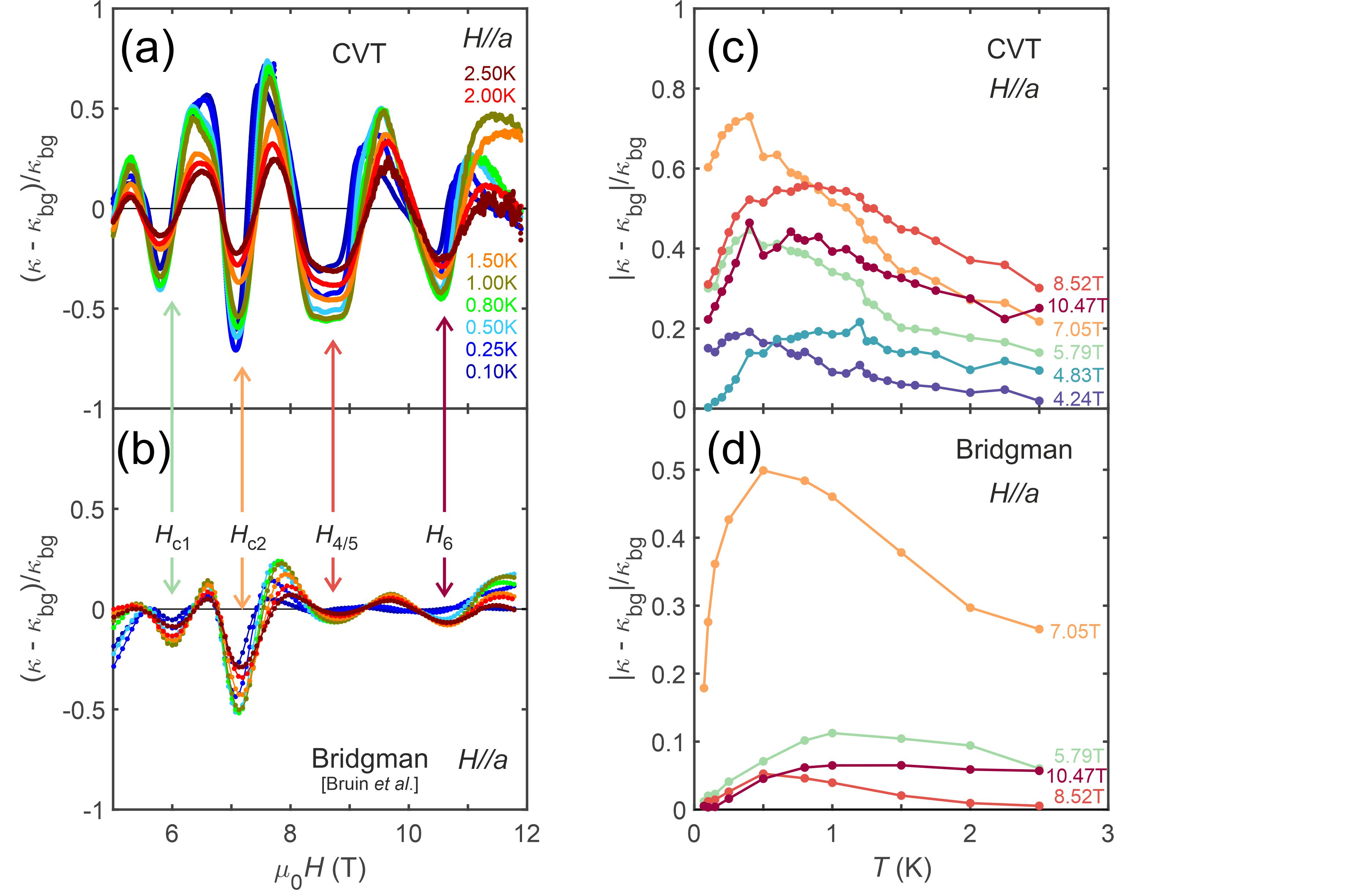}
\caption{\label{fig2} High-field oscillatory component of $\kappa(H)$ expressed as $(\kappa-\kappa_{\text{bg}})/\kappa_{\text{bg}}$. (a,b) Field dependences for the CVT and Bridgman-grown samples for temperatures ranging from 0.1 K (purple) to 2.5 K (red). Vertical arrows indicate the well-defined minima at $H_{\text{C1}}$ = 5.79 T (green) and $H_{\text{C2}}$ = 7.05 T (orange) as well as two higher field minima arising from transitions at $H_{\text{4/5}}$ $\approx$ 8.52 T and $H_{\text{6}}$ = 10.47 T (red, purple), which all coincide with peaks or shoulders in d$M$/d$H$. Whilst the structure at $H_{\text{4/5}}$ appears to consist of two minima in the CVT sample, only one broad minimum can be identified in the Bridgman sample. Underlying $\kappa(H)$ data for the Bridgman sample were reproduced from Ref.~\onlinecite{Bruin2022}. (c,d) Temperature dependences of the amplitudes of the minima in $(\kappa-\kappa_{\text{bg}})/\kappa_{\text{bg}}$ for the CVT and Bridgman samples for all fields where minima can be identified. Upon cooling from 2.5 K to $\sim$1 K, the amplitudes grow, but they rapidly collapse upon further cooling below 1 K.}
\end{figure*}

For both the Bridgman and the CVT crystals, the amplitudes of the oscillatory structures initially grow upon cooling, but then turn down sharply at lowest temperatures below 1 K. We demonstrate this by first subtracting a background (non-oscillatory) magnetic field dependence $\kappa_{\text{bg}}$ at each temperature by fitting a spline through the locations of greatest field derivative in $\kappa$. By plotting the isotherms of the normalized oscillatory content $(\kappa-\kappa_{\text{bg}})/\kappa_{\text{bg}}$ for $H \parallel a$ in Figs.~\ref{fig2}a and \ref{fig2}b, it is apparent that the amplitudes do not increase monotonically upon cooling, but instead decrease for lowest temperature, especially those of the highest field features. The temperature dependences of the amplitudes at the minima fields (Figs.~\ref{fig2}c and \ref{fig2}d) show this turnover between 500 mK - 1 K, which is observed in both samples. We note that the present data are consistent with those reported by Czajka \textit{et al.} \cite{Czajka2021a} which were taken above $\sim$1 K, and we only observe the decrease by further lowering temperature. Our observation speaks against Lifshitz-Kosevich behavior and a quantum oscillation interpretation.

What is then the origin of the structures in the magnetothermal conductivity? We previously argued\cite{Bruin2022} that the high-field features observed in the Bridgman sample may be due to magnetic phase transitions or crossovers, as two minima in $\kappa(H)$ coincide with the magnetic phase transitions at $H_{\text{C1}}$ and $H_{\text{C2}}$. The minima would then be caused by soft magnetic scattering of phonons, which dominate the thermal conductivity in $\alpha$-RuCl$_3$ \cite{Hentrich2020}. We argue that the high-field transitions at $H_{\text{4}}$ and $H_{\text{5}}$ are connected to weak magnetic transitions which have higher transition temperatures than the 7 K main phase. In zero field, signatures of transitions around 13 K \cite{Johnson2015,He2018,Mi2021} and 10 K \cite{Sears2015,Kubota2015,He2018,Mi2021} were reported previously in the magnetic susceptibility, the specific heat and the dielectric constant, which were discussed to originate from secondary phases associated with stacking faults. As shown in Figs.~\ref{fig4}a and \ref{fig4}b, the 7 K, 10 K and 13 K transitions can be identified as sequential kink-like structures in the temperature dependence of magnetization at a low magnetic field of 1 T. 

\begin{figure*}
\includegraphics[width=1\textwidth, angle=0]{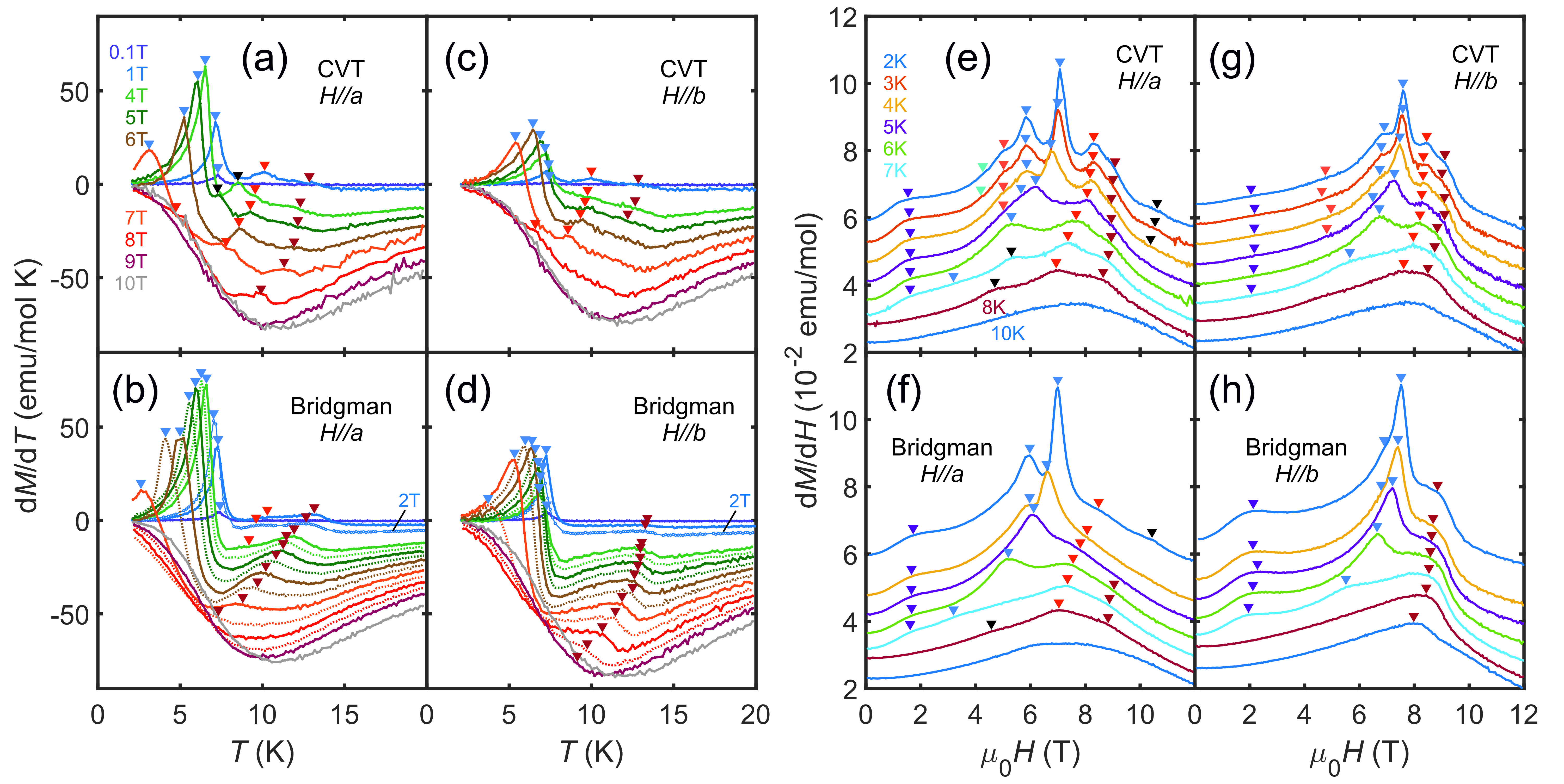}
\caption{\label{fig3} Tracking phase transitions in features of the temperature and field derivatives of magnetization of both the CVT and Bridgman samples. (a-d) d$M$/d$T$ at constant magnetic fields between 0.1 T and 10 T for $H \parallel a$ and $H \parallel b$. Colored triangles indicate the features tracking the 7 K (blue), 10 K (red) and 13 K (brown) transitions. Peaks of unknown origin are marked by black triangles. Dashed lines (Bridgman sample) show additional data taken at half-tesla intervals. (e-h) d$M$/d$H$ isotherms between 2 K and 10 K for $H \parallel a$ and $H \parallel b$ (shifted vertically for clarity). Blue triangles indicate the features tracking the main phase transitions at $H_{\text{C1}}$ and $H_{\text{C2}}$, additional colored triangles point at secondary features.}
\end{figure*}

The magnetic transitions in $M(T)$ can be identified more clearly as peaks in the temperature derivative d$M$/d$T$ as shown in Figs.~\ref{fig3}a-d, from which we can trace the magnetic field dependence of the phase boundaries. For the 7 K transition (dark blue triangles), the peak in d$M$/d$T$ becomes hard to define for fields above 6 T due to the rapid decrease of the transition temperature. Instead, well-defined peaks in d$M$/d$H$ appear at low temperature. They can be traced up to about 7 K (Figs.~\ref{fig3}e-h) and overlap with the transition temperature determined by the anomalies in d$M$/d$T$. Taken together, the peaks in d$M$/d$T$ and d$M$/d$H$ can be used to construct the entire phase diagram for a given direction of the in-plane magnetic field, including for features independent of the 7 K transition, as is shown in Figs.~\ref{fig4}c-f. We find that for both $H \parallel a$ and $H \parallel b$ the 10 K transition connects to $H_{\text{4}}$ for the CVT sample and the 13 K transition connects to $H_{\text{4/5}}$ for the Bridgman sample. The structures at the 13 K transition for the CVT sample and the 10 K transition for the Bridgman sample are difficult to track below $\sim$10 K but appear to connect to the $H_{\text{4/5}}$ features, although detailed behavior is complex and not fully traceable.

\begin{figure*}
\includegraphics[width=1\textwidth, angle=0]{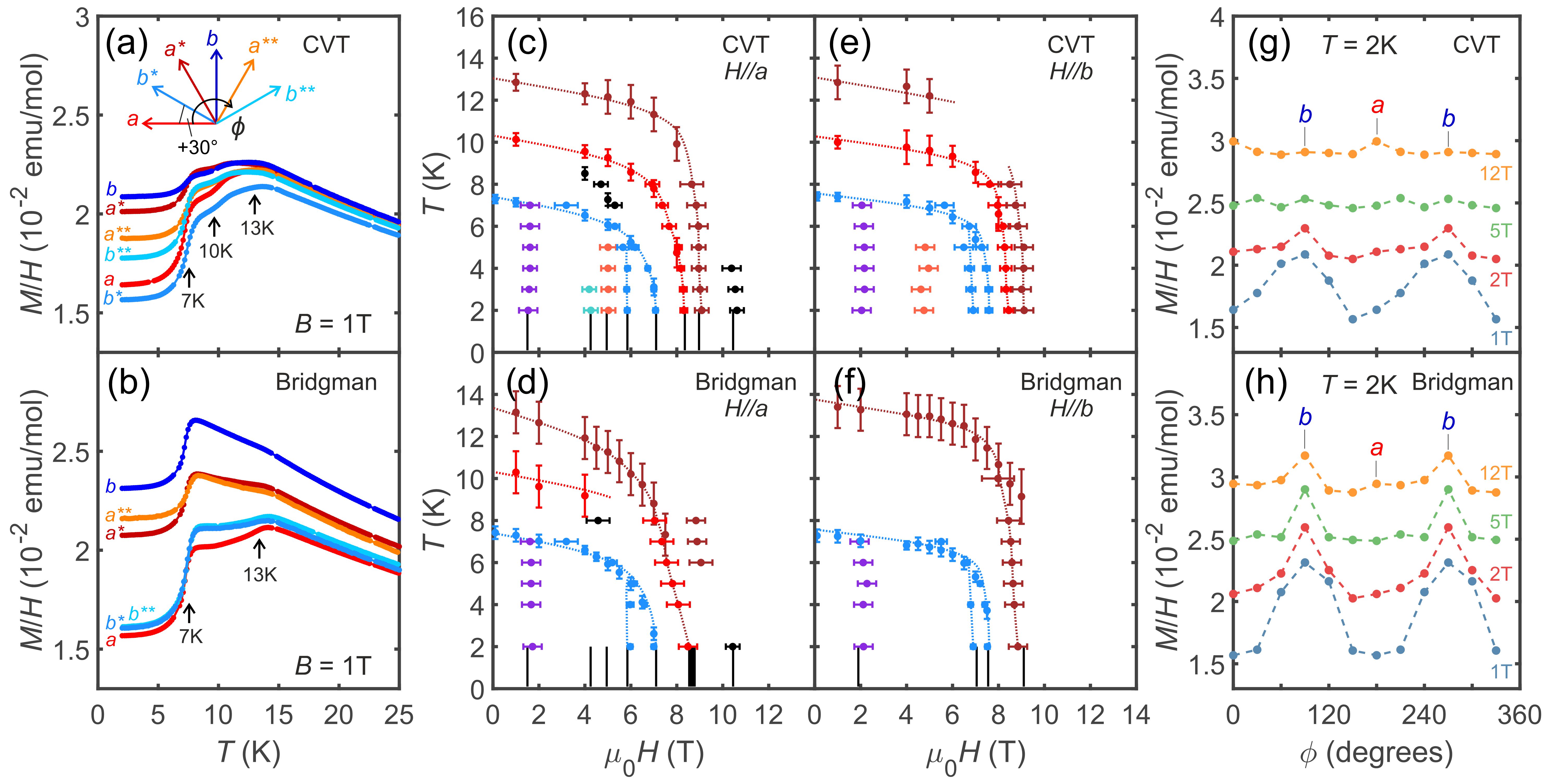}
\caption{\label{fig4} (a,b) Temperature dependence of $M/H$ at $B=1\textrm{T}$ for the CVT and Bridgman crystals, respectively. The six different magnetic field orientations, separated by $30^{\circ}$ rotations in the honeycomb plane are sketched in the inset of (a). Kink-like features at the three phase transitions at 7 K, 10 K and 13 K are indicated with arrows. (c-f) Phase diagrams for the CVT and Bridgman crystals with magnetic field along the $a$ and $b$-axes, constructed from the extracted maxima in d$M$/d$H$ (horizontal error bars) and d$M$/d$T$ (vertical error bars) (see: Fig.~\ref{fig3}). The sizes of the error bars represent the widths of the measured features. Solid black lines at low temperature trace the measured locations of minima in $\kappa(H)$ (see: Fig.~\ref{fig1}). Dashed lines are guides to the eye tracing possible phase transition lines. (g,h) $M/H$ at $T=2\textrm{K}$ as a function of in-plane-field angle $\phi$ at different magnetic fields, showing the gradual suppression of two-fold modulation upon increasing field.}
\end{figure*}

We conclude that the dip structures in $\kappa(H)$ at $H_{\text{4}}$ and $H_{\text{5}}$ indicated by the black lines at low temperatures in Figs.~\ref{fig4}c-f coincide reasonably with the phase boundaries connecting to the 10 K and 13 K transitions in zero field, just like the connection between $H_{\text{C1}}$ and $H_{\text{C2}}$ and the main 7 K transition, and therefore originate from the high temperature magnetic transitions. The 13 K and 10 K transitions have been discussed to originate from secondary phases closely related to stacking faults sensitive to mechanical deformation \cite{Cao2016,He2018,Mi2021,Banerjee2016}. The crystals used in this study have a soft, foil-like morphology and are easily deformed. Mechanical bending of the CVT crystal indeed enhances the structure at $H_{\text{4}}$ and $H_{\text{5}}$ in d$M$/d$H$ as shown in Fig.~\ref{fig1}a, which can be ascribed to the increase of volume of the secondary 10 K and 13 K phases by bending and support these phases as the origin of structure in $\kappa(H)$ and $M(H)$.

\begin{figure*}
\includegraphics[width=0.8\textwidth, angle=0]{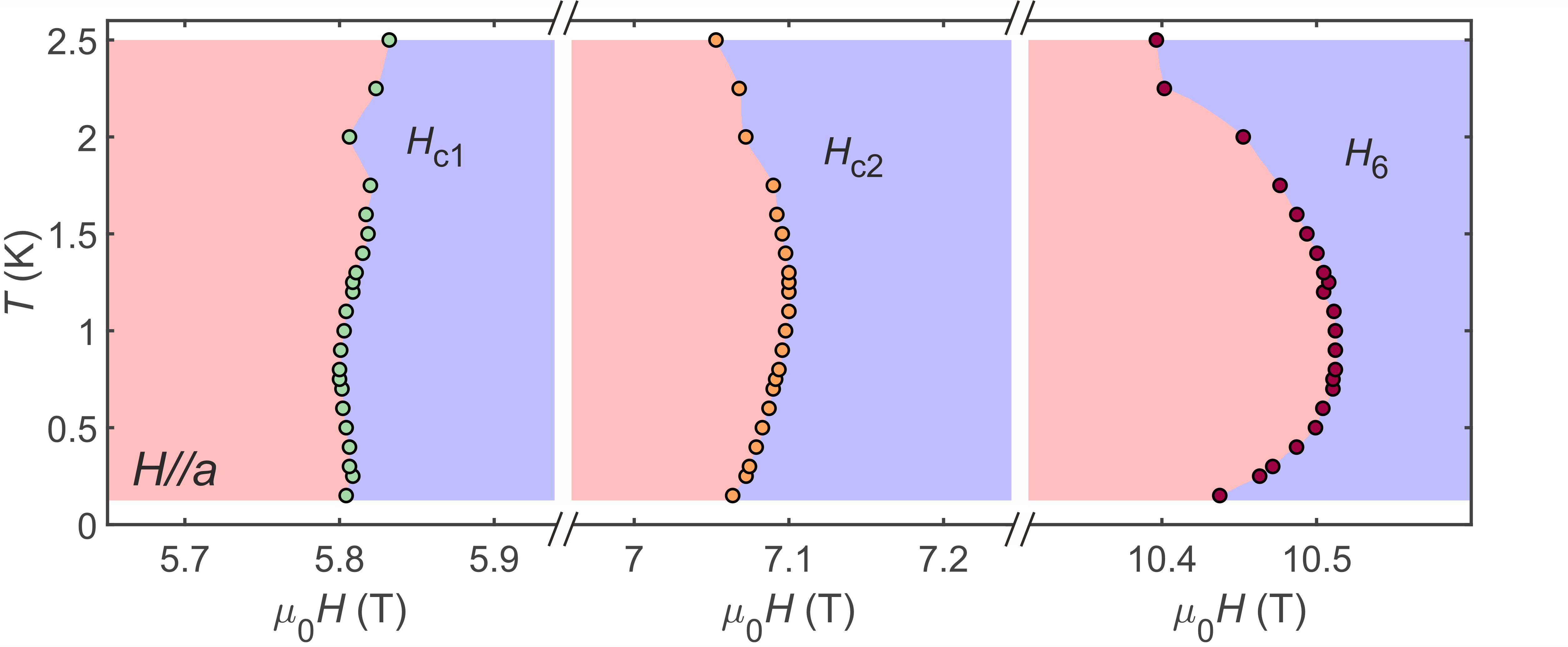}
\caption{\label{fig5} Locations of minima in $\kappa(H)$ of the CVT sample at $H_{\text{C1}}$, $H_{\text{C2}}$ and $H_{\text{6}}$ on the $H$-$T$ plane. Background colors indicate regions of negative d$\kappa(H)$/d$H$ (pink) and positive d$\kappa(H)$/d$H$ (blue) which define the locations of the minima in $\kappa(H)$ at their boundary. Field scales are expanded to emphasize the convex nature and inverse melting displayed by the transition lines of $H_{\text{C2}}$ and $H_{\text{6}}$.}
\end{figure*}

In Fig.~\ref{fig1}a, we see a bending-induced enhancement of the structures at $H_{\text{3}}$ and $H_{\text{6}}$ similar to those at $H_{\text{4}}$ and $H_{\text{5}}$. It strongly suggests that the structures at $H_{\text{3}}$ and $H_{\text{6}}$ are also linked to the phase transitions of bending-induced secondary phases. As the features in d$M$/d$H$ at $H_{\text{3}}$ and $H_{\text{6}}$ are weakly resolved at 2 K and rapidly disappear upon heating above 4 K (Figs.~\ref{fig3}e,f), we cannot trace them to corresponding phase transitions at low fields. Instead, we find that additional evidence for the phase transition origin of all the high-field features above $H_{\text{C1}}$ comes from the small but appreciable temperature dependence of positions of high-field dips in the magnetothermal conductivity. These are not expected for canonical (single frequency) quantum oscillations and support magnetic phase transitions as the origin of the structures in $\kappa(H)$. The phase transition lines of $H_{\text{C1}}$, $H_{\text{C2}}$ and $H_{\text{6}}$ are traced Fig.~\ref{fig5}. We also observed temperature dependences for $H_{\text{4}}$ and $H_{\text{5}}$ (see: Figs.~\ref{fig2}a and \ref{fig2}b) but due to the flatness of the associated minimum in $\kappa(H)$ the assignment of exact fields to those two features is less certain. The transition line of $H_{\text{C2}}$ (7 K main phase) as defined by minima in $\kappa(H)$ has a convex shape below 2.5 K with a field width of $\sim$70 mT, which agrees very well with the shape of that phase transition determined by a recent measurement of the Gr\"uneisen parameter, where it was discussed as an \textit{inverse melting} due to the higher entropy of the low-field ordered phase at very low temperatures \cite{Bachus2021}. The similarity in behavior of the features at $H_{\text{C2}}$ and $H_{\text{6}}$ confirms that the latter is also a thermodynamic phase transition with an appreciable entropy effect even below 1 K. Suetsugu \textit{et al.}\cite{Suetsugu2022} recently reported a rapid increase in $\kappa(H)$ around 11 T, where we observe the $H_{\text{6}}$ anomaly, and discussed that it represents a first-order topological transition from a quantum spin liquid with a half-quantized thermal Hall effect to a topologically trivial phase. Although the scenario proposed by Suetsugu \textit{et al.} is attractive and cannot be excluded by the present data, we note that the similarity between the features at $H_{\text{6}}$ and $H_{\text{C2}}$ in $\kappa(H)$, d$M$/d$H$ and in their inverse melting behavior means that the origin of $H_{\text{6}}$ is likely a magnetic phase transition in a secondary phase related to stacking faults. 

Finally, we suggest that the feature at $H_{\text{1}}$, and possibly also that at $H_{\text{2}}$, might originate from pseudo-spin reorientations within the zig-zag ordered phase, which is supported by the evolution of the in-plane field angle ($\phi$) dependence of magnetization in this field range. At 1 T (below $H_{\text{1}}$) $M/H$ in Figs.~\ref{fig4}a and \ref{fig4}b shows a clear two-fold anisotropy implying an easy-axis fixed along $a$, consistent with the reported $ac$-plane orientation of pseudospins in the zig-zag phase \cite{Cao2016}. The observed easy-axis anisotropy along $a$ is consistent with the monoclinic structure of the Bridgman sample, but in the trigonal CVT sample the origin of the anisotropy is unclear, perhaps a weak easy-axis anisotropy could be induced by strain, or alternatively the low temperature crystal structure may have a lowered symmetry due to the reported structural phase transition around 150 K \cite{He2018,Mi2021,Park2016}. The antiferromagnetic easy axis will rotate to perpendicular to the field direction when the field is strong enough to overcome the lattice pinning, resulting in a spin-flop transition. Such behavior was previously reported in studies of the magneto-optical dichroism \cite{Wagner2022} and terahertz spectroscopy \cite{Wu2018} at $\sim$1.5 T, where we observe the anomaly at $H_{\text{1}}$. Increasing the in-plane field from 1 T to 5 T suppresses the easy-axis anisotropy fully in the CVT crystal and substantially in the Bridgman crystal, as can be seen in Figs.~\ref{fig4}g and \ref{fig4}h. It is therefore likely that spin-flop or reorientation transitions have happened in between.

\section{Conclusion}
In summary, we show that by measuring the low temperature features in the thermal conductivity and magnetization of two $\alpha$-RuCl$_3$ single crystals grown with different sample growth techniques, a consistent picture emerges to explain high-field features. Minima in $\kappa(H)$ coincide with peaks in d$M$/d$H$, tracing out the magnetic phase transitions of both the 7 K main phase and other likely coexistent secondary phases, including those which have transition temperatures of 10 K and 13 K. These phases are enhanced by sample manipulation and exist in both types of samples. It is interesting to consider that whilst we treat them as essentially independent of the 7 K main phase and coexistent with it, it is clear that they strongly affect the bulk thermal transport, and their existence was shown to suppress the thermal Hall effect rapidly at low temperatures \cite{Bruin2022}. It would be highly desirable to isolate samples with only one of the 7 K, 10 K or 13 K phases present, but whether this is feasible in bulk crystals is still an open question.

\begin{acknowledgments}
We thank Y. Kasahara, Y. Matsuda, S. Suetsugu and H. Suzuki for insightful discussions and M. Dueller and K. Pflaum for technical assistance. The work done in Germany has been supported in part by the Alexander von Humboldt foundation. H.TAN and N.K. have been supported by JSPS KAKENHI Grant Number JP17H01142 and JP19K03711, respectively. H.TAK has been supported in part by JSPS KAKENHI, Grant Numbers JP22H01180 and JP17H01140. S.L. thanks the Science and Engineering Research Board (SERB), Government of India, for the award of a Ramanujan Fellowship (RJF/2021/000050).
\end{acknowledgments}

\section*{Author Declarations}
\subsection*{Conflict of Interest}
The authors have no conflicts to disclose.

\section*{Data Availability}
The data that support the findings of this study are available from the corresponding author upon reasonable request.


\bibliography{RuCl3collection}

\end{document}